\documentclass[a4paper,numcites,sort&compress,final]{aipproc}
\setcitestyle{numbers}
\layoutstyle{8x11double}

\usepackage{amsmath,amssymb} 

\usepackage{graphicx}

\usepackage{bm}
\DeclareMathOperator{\Tr}{Tr}

\newcommand{\beq}{\begin{eqnarray}}
\newcommand{\eeq}{\end{eqnarray}}

\renewcommand{\o}{\omega}

\newcommand{\vk}{\vec{k}}
\newcommand{\vq}{\vec{q}}
\newcommand{\vD}{\vec{D}}
\newcommand{\pli}{\prod\limits}
\newcommand{\rk}{\right)}
\newcommand{\lk}{\left(}
\newcommand*{\be}{\begin{equation}}
\newcommand*{\ee}{\end{equation}}

\newcommand*{\vev}[1]{\left< #1 \right>}
\newcommand*{\Eqref}[1]{Eq.\ \eqref{#1}}
\newcommand*{\abs}[1]{\ensuremath{\lvert#1\rvert}}
\renewcommand*{\i}{\ensuremath{\mathrm{i}}}

\newcommand*{\vp}{{\vec{p}}}
\providecommand*{\coloneq}{\mathrel{\mathop:}=}

\newcommand{\N}{{\cal N}}

\newcommand{\oh}{\frac{1}{2}}

\newcommand{\non}{\nonumber}

\newcommand{\ra}{\rightarrow}

\renewcommand*{\vec}[1]{\bm{#1}}
\begin{document}

\title[]{Hamiltonian approach to Yang-Mills theory in Coulomb gauge - revisited${}^{\P}$}

\classification{}
\keywords{}

\author{H.~Reinhardt}%
{address={Institut f\"ur Theoretische Physik, Universit\"at T\"ubingen, Auf der Morgenstelle 14, 72076 T\"ubingen, Germany}}

\author{D.~R.~Campagnari}%
{address={Institut f\"ur Theoretische Physik, Universit\"at T\"ubingen, Auf der Morgenstelle 14, 72076 T\"ubingen, Germany}}

\author{M.~Leder}%
{address={Institut f\"ur Theoretische Physik, Universit\"at T\"ubingen, Auf der Morgenstelle 14, 72076 T\"ubingen, Germany}}

\author{G.~Burgio}%
{address={Institut f\"ur Theoretische Physik, Universit\"at T\"ubingen, Auf der Morgenstelle 14, 72076 T\"ubingen, Germany}}

\author{J.~M.~Pawlowski}%
{address={Institut f\"ur Theoretische Physik, Universit\"at Heidelberg, Philosophenweg 16, 69120 Heidelberg, Germany}}

\author{M.~Quandt}%
{address={Institut f\"ur Theoretische Physik, Universit\"at T\"ubingen, Auf der Morgenstelle 14, 72076 T\"ubingen, Germany}}

\author{A.~Weber}%
{address={Instituto de F\'\i{}sica y Matem\'aticas, Universidad Michoacana de San Nichol\'as de Hidalgo, Edificio C-3,
Ciudad Universitaria, 58040 Morelia, Michoac\'an, Mexico}}

%\iftrue
%\author{Arno Mittelbach}{
%  address={Zedernweg 62, 55128 Mainz, Germany},
%  email={arno@mittelbach-online.de},
%}

%\author{D. P. Carlisle}{
%  address={Willow House, Souldern},
%  email={david@dcarlisle.demon.co.uk},
%  homepage={http://www.dcarlisle.demon.co.uk},
%  altaddress={When I go to work: NAG Ltd, Oxford}
%}
%\fi

% \copyrightholder{Acoustical Scociety of America}
%\copyrightyear  {}

\begin{abstract}
I briefly review results obtained within the variational Hamiltonian approach to Yang-Mills theory in Coulomb gauge and confront them with recent lattice data. The variational approach is extended to non-Gaussian wave functionals including three- and four-gluon kernels in the exponential of the vacuum wave functional and used to calculate the three-gluon vertex. A new functional renormalization group flow equation for Hamiltonian Yang--Mills theory in Coulomb gauge is solved for the gluon and ghost propagator under the assumption of ghost dominance. The results are compared to those obtained in the variational approach.
\end{abstract}

\date{\today}

\maketitle

\section{Introduction}

In recent years there have been substantial efforts devoted to non-perturbative studies
of continuum Yang--Mills theory. Among these are variational solutions of the Yang--Mills
Schr\"odinger equation in Coulomb gauge \citep{SzcSwa01,FeuRei04,EppReiSch07}. In this
approach the energy density is minimized using Gaussian-type ans\"atze for the vacuum
wave functional. In this talk I will first briefly review results obtained within the
T\"ubingen approach \citep{FeuRei04} using a modified Gaussian wave functional. Then I
will present a new approach to Hamiltonian quantum field theory, which allows to use
non-Gaussian wave functionals in variational calculations \citep{CamRei10}. The approach
is illustrated by using a wave functional with cubic and quartic gluonic terms in the
exponential and applied to calculate the three-gluon vertex. Finally, I will discuss a
new functional renormalization group (FRG) approach to the Hamiltonian formulation of
Yang--Mills theory \citep{Led+10} and present results supporting the findings of the
variational approach \citep{FeuRei04,EppReiSch07}. 
\renewcommand{\thefootnote}{\fnsymbol{footnote}}
\footnotemark[0]
\footnotetext[6]{Invited talk given by H.~Reinhardt at ``T(r)opical QCD 2010'',
September~26--October~1, 2010, Cairns, Australia.}
\renewcommand{\thefootnote}{\arabic{footnote}}

%%%%%%%%%%%%%%%%%%%%%%%%%%%%%%%%%%%%%%%%%%%%%%%%%%%%%%%%%%%%%%%%%%%%%%%%%%%%%%%%%%%%%%%%%

\section{Variational Approach}

Instead of working with gauge invariant wave functionals, it is usually more convenient to fix the gauge.
A particularly convenient choice of gauge is the Coulomb gauge $\vec{\nabla}\cdot\vec{A} = 0$,
which allows an explicit resolution of Gauss' law, resulting in the gauge fixed
Yang--Mills Hamiltonian \citep{ChrLee80}
\begin{align}
\label{398-G1}
H_{\textsc{ym}} &= \frac{1}{2} \int d^D x \left( J^{-1} [\vec{A}] \,{\vec{\Pi}} J[\vec{A}] \,\vec{\Pi}
+ \vec{B}^2 \right) + H_c , \\
H_c & = \frac{g^2}{2} \int d^D (x,x') J^{-1}[\vec{A}] \rho^a(x) J[\vec{A}] F^{ab}(x, y) \rho^b(x) , \nonumber
\end{align}
where $\Pi^a (x) = \delta/i \delta A^a (x)$ is the canonical momentum
(electric field) operator and
\beq
\label{404}
J[\vec{A}] = \mathrm{Det} (- \vec{D} \vec{\nabla})
\eeq
is the Faddeev--Popov determinant. Furthermore
\beq
\label{409-G3}
\rho^a (x) = - f^{abc} \vec{A}^b \vec{\Pi}^c
\eeq
is the color charge of the gluons and
\beq
\label{414-G4}
F^{ab} (x, y) = \langle x, a \rvert (- \vec{D} \vec{\nabla})^{- 1}
\, (- \vec{\nabla}^2)\, (- \vec{D} \vec{\nabla})^{- 1} \lvert y, b \rangle
\eeq
is the so-called Coulomb kernel. In the presence of matter fields with color
charge density $\rho^a_m (x)$,
the gluon charge $\rho^a (x)$ in the Coulomb term
is replaced by the total charge $\rho^a (x) + \rho^a_m (x)$ and the vacuum
expectation value of $F^{ab} (x, y)$ acquires the meaning of the static non-Abelian
Coulomb potential, see eq.~(\ref{gapeq}) below.

The gauge fixed Hamiltonian eq.~(\ref{398-G1}) is highly non-local due to the Coulomb
kernel $F^{ab} (x, y)$,
eq.~(\ref{414-G4}), and due to the Faddeev--Popov determinant $J [\vec{A}]$, eq.~(\ref{404}).
In addition, the latter also occurs in the functional integration measure of
the scalar product of Coulomb gauge wave functionals
\beq
\label{419-G5}
\langle \psi_1 | O | \psi_2 \rangle = \int D A \, J [\vec{A}] \, \psi^*_1[\vec{A}] \, O \, \psi_2 [\vec{A}] .
\eeq
While the elimination of unphysical degrees of freedom via gauge fixing is
often beneficial (and sometimes unavoidable) in practical calculations,
the price to pay is the increased complexity of the gauge fixed  Hamiltonian eq.~(\ref{398-G1}). The non-trivial Faddeev--Popov determinant reflects the intrinsically non-linear structure of the space of gauge orbits and dominates the IR behavior of the theory. Once Coulomb gauge is implemented, any functional of the (transverse) gauge field is a physical state. 

Treating the Hamiltonian (\ref{398-G1}) in the ordinary Rayleigh-Schr\"odinger perturbation theory one finds to leading order the usual one-loop $\beta$ function with $\beta_0 = - \frac{11}{3} N_c$ \citep{CamReiWeb09}. Our main interest lies, however, in the IR sector of the theory, which requires a non-perturbative treatment.

In ref. \citep{FeuRei04} the vacuum energy density was minimized using the following ansatz for the vacuum functional
\be
\label{412}
\Psi [\vec{A}]  = \frac{\N}{\sqrt{J[\vec{A}]}} \,
\exp\left[-\oh\int \frac{d^3k}{(2\pi)^3} \omega(\vec{k}) A_i^a(\vec{k}) A_i^a(-\vec{k}) \right] .
\ee
For this wave functional the static gluon propagator is given by\footnote{Here and in the following,
$t_{ij}(\vec{k}) \equiv \delta_{ij} - k_i\,k_j/\vec{k}^2$ denotes the transverse
projector in momentum space.}
\beq
\langle A^a_i (- \vec{k}) A^b_j (\vec{q}) \rangle = \delta^{ab} t_{ij}
(\vec{k}) (2 \pi)^3 \delta (\vec{k} - \vec{q}) \frac{1}{2 \omega (\vec{k})}\,,
\eeq
so that $\omega (\vec{k})$ represents the (quasi-) gluon energy.

To the order considered in ref.~\citep{FeuRei04}, i.e., up to two loops
in the energy (which corresponds to one loop in the associated Dyson--Schwinger
equations), it was shown \citep{ReiFeu05} that the Faddeev--Popov
determinant, eq.~(\ref{404}), can be represented as
\beq
\label{430}
    J[\vec{A}] = \exp\left[-\int d^3 x \, d^3 y \: A^a_i (x) \chi^{ab}_{ij} (x, y) A^b_j (y)  \right] ,
\eeq
where
\be
\label{455-38}
\chi^{ab}_{ij} (x, y) = - \frac{1}{2} \left\langle \frac{\delta^2 \ln J[\vec{A}]}{\delta A^a_i (x) \delta A^b_j (y) } \right\rangle =
\delta^{ab} t_{ij} (x)  \chi (x, y)
\ee
is the ghost loop and represents a measure of the {\it curvature} \citep{FeuRei04} of the
Coulomb gauge fixed configuration space. Expressing the ghost Green's function as
\beq
\langle (- D \vec{\nabla})^{- 1} \rangle = \frac{d}{g (- \Delta)} \, ,
\eeq
where $d (\vec{k})$ is the  ghost form factor, its Dyson--Schwin\-ger equation reads to one-loop
order
\begin{gather}
\label{G45}
d^{-1}(\vec{k}) = g^{-1} - I_d(\vec{k}) , \displaybreak[1]\\
I_d(\vec{k}) = \frac{N_c}{2} \int \frac{d^3 q}{(2 \pi)^2}\, \left[1 - (\hat{\vk} \cdot
\hat{\vq})^2\right]\, \frac{d(\vec{q}-\vec{k})}{ \o(\vec{q})\, (\vec{q}-\vec{k})^2} \, .
\end{gather}
The ghost form factor $d(\vec{k})$ measures the deviation from the QED case, where
$d(\vec{k})=1$. Furthermore, $d^{- 1} (\vec{k})$ represents the dielectric function of
the Yang--Mills vacuum, and the so-called horizon condition
\be
\label{313-xx}
d^{- 1} (0) = 0
\ee
ensures that the Yang--Mills vacuum is a dual superconductor \citep{Rei08}. 
Minimization of the vacuum energy $\langle H_\textsc{ym} \rangle$ with respect to
the kernel $\omega (\vec{k})$
leads to the gap equation, which after renormalization reads \citep{FeuRei04}, \citep{ReiEpp07}
\begin{multline}
\label{G49}
\o^2(\vec{k}) = \vec{k}^2 + \chi^2(\vec{k}) + c_2 + \Delta I^{(2)}(\vec{k})  + \\ + 2 \chi(\vec{k})
\,[\Delta I^{(1)}(\vec{k}) + c_1] \, ,
\end{multline}
where
\begin{align}
 \Delta I^{(n)}(\vec{k}) ={}& 
 I^{(n)}(\vec{k}) - I^{(n)}(0) , \qquad \overline \omega(\vec{k}) \equiv \o(\vec{k}) - \chi(\vec{k}) , \nonumber \\
I^{(n)}(\vec{k}) 
={}& \frac{N_c}{4} \int \frac{d^3 q}{(2 \pi)^3} \lk 1 +(\hat{\vk}\cdot\hat{\vq})^2 \rk V(\vec{q}-\vec{k}) \non \\
&\qquad\qquad \times \frac{ \overline{\o}^n(\vec{q}) - \overline{\o}^n(\vec{k})}{\o(\vec{q})} \, ,
\end{align}
and
\be\label{gapeq}
V(\vec{k}) \equiv g^2 \,\langle F \rangle = \frac{d^2 (\vec{k})}{\vec{k}^2}
\ee
is the non-Abelian Coulomb potential, where the approximation
\begin{multline}
\label{512}
\big\langle\, (- \vD \vec{\nabla})^{- 1}\cdot (- \vec{\nabla}^2) \cdot(- \vD
\vec{\nabla})^{- 1} \,\big\rangle \\
\qquad\qquad{}\approx\big\langle\, (- \vD \vec{\nabla})^{- 1}\,\big\rangle
\cdot(- \vec{\nabla}^2) \cdot\big\langle\, (- \vD \vec{\nabla})^{- 1} \,\big\rangle
\end{multline}
has been used.
In the gap equation (\ref{G49}), $c_1$ and $c_2$ are (finite) renormalization constants.
For the critical solution, where one imposes the horizon condition (\ref{313-xx}),  both
$\o(\vec{k})$ and $\chi(\vec{k})$ are infrared divergent,  which implies that the
transverse gluon propagator vanishes at $k \ra 0$, while
\beq
\overline \o(0) \equiv \lim_{k \ra 0} (\o(\vec{k}) - \chi(\vec{k})) = c_1.
\label{renormc1}
\eeq
A perimeter law of the 't~Hooft loop  requires $c_1 = 0$
and this value is also favored by the variational
principle \citep{ReiEpp07}.  Furthermore, for $c_1 = 0$ the IR limit
of the wave functional (\ref{412}) is
\beq
\label{532}
\psi [\vec{A}] = \N \pli_{\vec{k}} \psi (\vec{k}) , \qquad \psi (\vec{k} = 0) = 1 ,
\eeq
which is the exact vacuum wave functional in $1 + 1$ dimensions
\citep{ReiSch08}. The renormalization parameter $c_2$, on the other hand,
has no influence on the IR or UV behavior of the solutions of the gap equation
(\ref{G49}). Only the mid-momentum regime of $\omega (\vec{k})$ is weakly dependent
on $c_2$ \citep{FeuRei04}. Since we are mainly interested in the IR
properties we will put $c_2 = 0$.

\begin{figure}[t]
\includegraphics[width=0.85\linewidth]{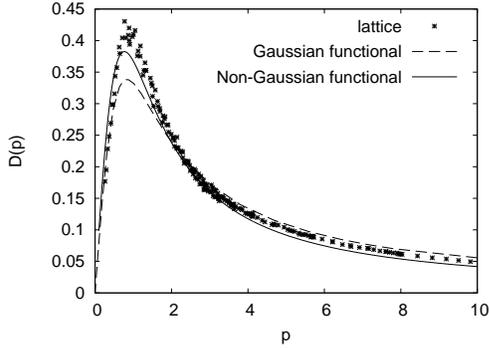}
\caption{\label{fig:prop}Gluon propagator obtained with a Gaussian (dashed line) and a
non-Gaussian functional (straight line), compared to the lattice data from Ref.\
\citep{BurQuaRei09}.}
\end{figure}
\begin{figure}[t]
\centerline{\includegraphics[width=.85\linewidth]{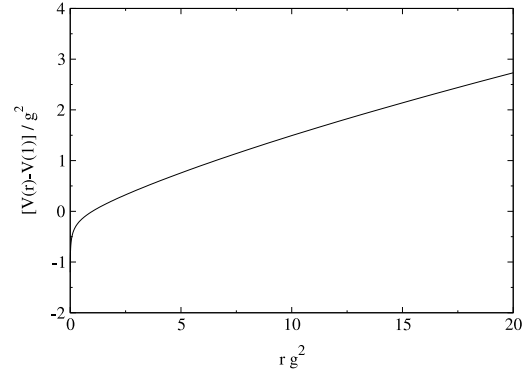}}
\caption{The numerical solution for the static non-Abelian Coulomb
potential~\citep{EppReiSch07}.}
\label{tuefig3}
\end{figure}

The coupled integral equations \eqref{G45} and \eqref{G49} can be solved analytically
in the IR (for the critical solution satisfying the horizon condition (\ref{313-xx}))  using power law ans\"atze
\citep{FeuRei04,SchLedRei06}
\beq
\label{444}
\omega (\vec{k}) \sim k^{- \alpha}, \qquad d (\vec{k}) \sim k^{- \beta} .
\eeq
Under the assumption of a trivial scaling of the ghost-gluon vertex
one finds that in $D$ space dimensions the  IR exponents satisfy the sum rule
\beq
\label{409}
\alpha = 2 \beta + 2 - D .
\eeq
This sum rule ensures that $\chi (\vec{k})$ has the same IR behavior as $\omega (\vec{k})$,
cf.~eq.~(\ref{renormc1}).
Furthermore, in $D = 3$ one finds two solutions \citep{SchLedRei06}
\be
\label{453}
\beta \simeq 0.8 \quad  \text{and} \quad  \beta = 1  .
\ee
The latter gives rise to a Coulomb potential (\ref{gapeq}) which is strictly
linear at large distances. In $D = 2$ only one solution $\beta = \frac{2}{5}$ is found,
implying $\alpha = \frac{4}{5}$ 
and a Coulomb potential (\ref{gapeq}) rising as
$V (r) \sim r^{4/5}$ \citep{FeuRei08}.
Furthermore in $D = 2$ only the critical solution satisfying (\ref{313-xx}) exists \citep{FeuRei08}.
On the lattice, on the other hand, one finds $\alpha=1$ in $D=3$ and a linearly rising potential
\citep{BurQuaRei10}.
In the UV-regime the coupled equations (\ref{G45}) and (\ref{G49}) can be solved
perturbatively and one finds at large momenta $k \gg 1$~\citep{FeuRei04}
\be
\label{565}
\omega(\vec{k}) \sim k , \qquad d(\vec{k}) \sim 1/\sqrt{\log(k)}.
\ee
The full numerical solutions of the above equations were given, for $D = 3$
space dimensions, in refs.~\citep{FeuRei04} and \citep{EppReiSch07}, where the
critical solutions  $\beta = 0.8$ and $\beta = 1$, respectively, were found, and
for $D = 2$ in ref.~\citep{FeuRei08}.
One finds an inverse gluon propagator which in the UV behaves like the
photon energy but diverges in the IR, signaling confinement in agreement with the IR analysis (\ref{444}), (\ref{409}), (\ref{453}). Figures \ref{fig:prop} and \ref{tuefig3} show the resulting gluon energy $\omega (\vec{k})$ and non-Abelian Coulomb potential (\ref{gapeq}) for the solution $\beta = 1$.
Their IR behavior is in agreement with the results of the IR analysis. The obtained
propagator also compares favorably with the available lattice data \citep{BurQuaRei09}. There are, however,
deviations in the mid-momentum regime (and minor ones in the UV) which can be attributed
to the missing gluon loop, which escapes the Gaussian wave functionals. These deviations
are presumably irrelevant for the confinement properties, which are dominated by the ghost
loop (which is fully included under the Gaussian ansatz), but are believed to be important
for a correct description of spontaneous breaking of chiral symmetry \citep{YamSug10}. 

%%%%%%%%%%%%%%%%%%%%%%%%%%%%%%%%%%%%%%%%%%%%%%%%%%%%%%%%%%%%%%%%%%%%%%%%%%%%%%%%%%%%%%%%%

\section{Non-Gaussian wave functionals}

In ref. \citep{CamRei10} the variational approach to Yang--Mills theory in Coulomb gauge was extended to non-Gaussian wave functionals of the form
\begin{gather}
\label{511-xx}
\lvert \psi[\vec{A}] \rvert^2 = \exp (- S[\vec{A}]) , \\
\label{516-xx}
S [A] = \int \omega \vec{A}^2 + \frac{1}{3!} \int \gamma_3 \vec{A}^3 + \frac{1}{4!} \int \gamma_4 \vec{A}^4 ,
\end{gather}
where $\omega$, $\gamma_3$, $\gamma_4$ are variational kernels.

Representing the Faddeev--Popov determinant by a functional integral over ghost fields from the identity
\be
\label{523-xx}
0 = \int D A \frac{\delta}{\delta A} \lk J[\vec{A}] \mathrm{e}^{- S [\vec{A}]} K [\vec{A}] \rk ,
\ee
with $K [\vec{A}]$ an arbitrary functional of the gauge field, one can derive, in the standard
fashion, DSEs for the various gluon and ghost Green functions. These DSEs are the usual
DSEs of Landau gauge Yang--Mills theory, however, in $D=3$ dimensions and with the bare
vertices of the usual Yang--Mills action replaced by the variational kernels $\omega$,
$\gamma_3$, $\gamma_4$. It should be stressed that these Hamiltonian DSEs are not
equations of motion in the usual sense, but rather relations between the Green functions
and the so far undetermined variational kernels.
By using these DSEs, the energy density can be written as a functional
of the variational kernels,
\be\label{mad5a}
\vev{H_\textsc{ym}} = E[\omega,\gamma_3, \gamma_4] .
\ee
By using a skeleton expansion, the vacuum energy can be written at the desired order of
loops. Confining ourselves to two loops, the variation of the vacuum energy \Eqref{mad5a}
with respect to the kernel $\gamma_3$ fixes the latter to
\begin{multline}\label{mad6}
\gamma^{abc}_{ijk}(\vp,\vq,\vk) = 2 \, \i \, g \, f^{abc} \\
\times \frac{ \delta_{ij} (p-q)_k + \delta_{jk} (q-k)_i +\delta_{ki} (k-p)_j }{\Omega(\vp) + \Omega(\vq) + \Omega(\vk)} .
\end{multline}
Equation \eqref{mad6} is reminiscent of the lowest-order perturbative result \citep{Cam+09},
with the perturbative gluon energy $\abs{\vp}$ replaced by the non-perturbative one
$\Omega(\vp)$.

With this result, variation of $\langle H_\textsc{ym}\rangle$ with respect to $\omega$
yields the gap equation (\ref{G49}), which now, however, contains on the r.h.s.\ in addition
the gluon loop $I_g (\vec{k})$ \citep{CamRei10}, which escapes the Gaussian wave functional (\ref{412}). 
The presence of the gluon loop $I_g(\vec{k})$ in the gap
equation modifies the UV behavior and allows us to extract, from the
non-renormalization of the ghost-gluon vertex, the correct first coefficient of the
$\beta$ function. In order to estimate the size of the gluon-loop contribution to
the gluon propagator, we use the gluon and ghost propagators obtained with a Gaussian
wave functional \citep{EppReiSch07} to calculate the gluon loop. The result is shown in
Fig.\ \ref{fig:prop}, together with lattice data from Ref.\ \citep{BurQuaRei09}.
The agreement between the continuum and the lattice results is improved in the
mid-momentum regime by the inclusion of the gluon loop, i.e., the three-gluon vertex,
as observed also in Landau gauge \citep{FisMaaPaw09}. 

\begin{figure}
\includegraphics[width=.6\linewidth]{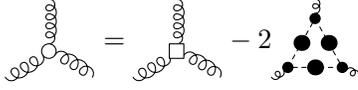}
\caption{\label{fig:3gvdse}Truncated DSE for the three-gluon vertex, under the assumption
of ghost dominance.}
\end{figure}

\begin{figure}
\includegraphics[width=0.9\linewidth]{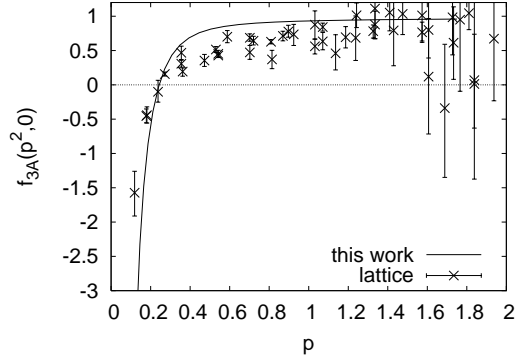}
\caption{\label{fig:3gv}Form factor $f_{3A}$ of the three-gluon vertex for orthogonal
momenta and comparison to lattice data for the $d=3$ Landau-gauge vertex \citep{CucMaaMen08}.
The momentum scale is arbitrary and has been adjusted to make the sign change occur at
the same point. The lattice data are shown by courtesy of A.~Maas.}
\end{figure}

The truncated DSE for the three-gluon vertex $\Gamma_3$ under the assumption of ghost
dominance is represented diagrammatically in Fig.\ \ref{fig:3gvdse}.
Possible tensor decompositions of the three-gluon vertex are given in Ref.\
\citep{BalChi80b}. For sake of illustration, we confine ourselves to
the form factor corresponding to the tensor structure of the bare three-gluon vertex
\be\label{mad3gv1}
f_{3A} \coloneq
\frac{\Gamma_3 \cdot \Gamma_3^{(0)}}{\Gamma_3^{(0)} \cdot \Gamma_3^{(0)}},
\ee
where $\Gamma_3^{(0)}$ is the perturbative vertex, given by \Eqref{mad6} with $\Omega(\vp)$
replaced by $\abs{\vp}$. Furthermore, we consider a particular kinematic configuration,
where two momenta have the same magnitude
\be\label{mad3gv2}
\vp_1^2 = \vp_2^2 = p^2 , \quad \vp_1\cdot\vp_2 = c p^2 .
\ee
To evaluate the form factor $f_{3A}(p^2,c)$, we use the ghost and gluon propagators
obtained with a Gaussian wave functional \citep{EppReiSch07} as input. The IR analysis
of the equation for $f_{3A}(p^2,c)$ [\Eqref{mad3gv1}] performed in Ref.\
\citep{SchLedRei06} shows that this form factor should have a power law
in the IR, with an exponent three times the one of the ghost dressing function; this is
confirmed by our numerical solution \citep{CamRei10}. The result for the scalar form
factor $f_{3A}$ for orthogonal momenta, $f(p^2,0)$, is shown in Fig.\ \ref{fig:3gv},
together with lattice results for $d=3$ Landau gauge Yang--Mills theory.
Our result and the lattice data compare favorably in the low-momentum regime. In
particular, in both studies, the sign change of the form factor occurs roughly at the
same momentum where the gluon propagator has its maximum. (The scale in Fig.\ \ref{fig:3gv}
is arbitrary.)

%%%%%%%%%%%%%%%%%%%%%%%%%%%%%%%%%%%%%%%%%%%%%%%%%%%%%%%%%%%%%%%%%%%%%%%%%%%%%%%%%%%%%%%%%

\section{Hamiltonian flow}

The advantage of the variational approach to the Hamiltonian formulation is its close connection
to physics. The price to pay is the apparent loss of manifest renormalization group invariance. 
Renormalization group invariance is naturally built-in in the
functional renormalization group approach to the Hamiltonian formulation
of Yang--Mills theory proposed in ref. \citep{Led+10}.

In the FRG approach the quantum theory of a field $\varphi$ is infrared regulated by adding the regulator
term \be
\label{12}
\Delta S_k [\varphi] = \frac{1}{2} \varphi \cdot R_k \cdot \varphi
\equiv \frac{1}{2} \int \varphi R_k \varphi
\ee
to the classical action, which in the Hamiltonian approach is given by the logarithm of
the wave functional (\ref{511-xx}). The regulator function
$R_k (p)$ is an effective momentum dependent mass with the
properties
\be\label{13}
\lim\limits_{p/k \to 0} R_k (p) > 0 , \qquad \lim\limits_{k/p \to 0} R_k (p) = 0 ,
\ee 
which ensures that $R_k (p)$ suppresses propagation of modes
with $p \lesssim k$ while those with $p \gtrsim k$ are unaffected and
the full theory at hand is recovered as the cut-off scale $k$ is
pushed to zero. Wetterich's flow equation for the effective action $\Gamma_k[\phi]$ of a field $\phi$ is given by
\begin{gather}
\label{18}
k \partial_k \Gamma_k [\phi] = \frac{1}{2} \Tr\, 
\frac{1}{ \Gamma_k^{(2)} [\phi] + R_k } \, k \partial_k R_k  , \\
\label{19}
\Gamma^{(n)}_{k, 1 \dots n} [\phi] = \frac{\delta^n 
\Gamma_k[\phi]}{\delta \phi_1 \dots \delta \phi_n}
\end{gather}
are the one-particle irreducible $n$-point functions (proper
vertices). The generic structure of the flow equation (\ref{18}) is
independent of the details of the underlying theory, but is a mere consequence of
the form of the regulator term (\ref{12}), being quadratic
in the field. By taking functional derivatives of Eq. (\ref{18}) one
obtains the flow equations for the (inverse) propagators and proper
vertices. 
The general flow equation (\ref{18}) still holds for the Hamiltonian formulation of Yang--Mills theory provided that
$\phi$ is interpreted as the superfield $\phi = (\vec{A}, c, \bar{c})$. 

\begin{figure}[t]
\includegraphics[width=0.8\linewidth]{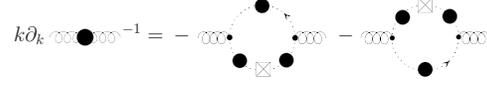}
\caption{Truncated flow equation of the gluon propagator. The bare vertices at $k=\Lambda$ are symbolized by small dots
and the regulator insertion $k \partial_k R$ by a crossed square.} 
\label{trunc flow gluon}
\end{figure}

\begin{figure}[t]
\includegraphics[width=0.8\linewidth]{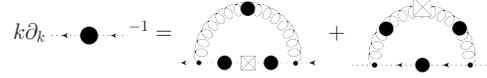}
\caption{Truncated flow equation of the ghost propagator} 
\label{trunc flow ghost}
\end{figure}

The FRG flow equations embody an infinite tower of coupled equations
for the flow of the propagators and the proper vertices. These
equations have to be truncated to get a closed system. We shall use
the following truncation: we only keep the gluon and ghost
propagators, which we write in the form
\be
\label{25}
\Gamma^{(2)}_{k,AA}  =  2 \omega_k (\vec{p}) , \qquad
\Gamma^{(2)}_{k,\bar{c} c}  =  \frac{p^2}{d_k (\vec{p})} \, .
\ee
In addition, we keep the ghost-gluon vertex $\Gamma^{(3)}_{k,A \bar{c} c}$, which we
assume to be bare, i.e., we do not solve its
FRG flow equation. Moreover, we shall assume infrared ghost
dominance and discard gluon loops. Then the resulting flow equations of the
ghost and gluon propagator reduce to the ones shown in Figs.~\ref{trunc flow gluon} and \ref{trunc flow ghost}.

\begin{figure}
\includegraphics[width=\linewidth]{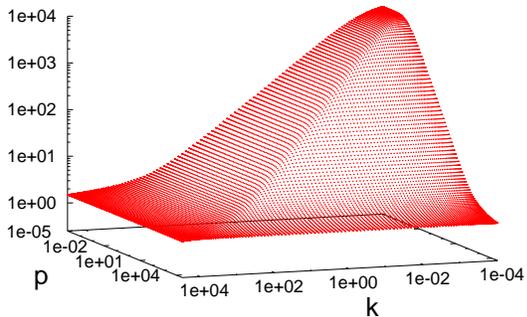}
\caption{Flow $d_k(p)$ of the ghost form factor.}
\label{ghost_flow_full.eps}
\end{figure}

These flow equations are solved numerically using the regulators
\begin{gather}
 R_{A,k} (\vec{p}) = 2p r_k (\vec{p}) , \qquad  R_{c,k} (\vec{p}) = p^2 r_k (\vec{p}) , \\
 r_k (\vec{p}) = \exp \left[ \frac{k^2}{p^2} - \frac{p^2}{k^2} \right] ,
\end{gather}
and the perturbative initial conditions at the large momentum scale $k = \Lambda$,
\be
\label{29}
d_{\Lambda} (p) = d_\Lambda = \mathrm{const.} \, , \qquad \omega_\Lambda (p) = p + a .
\ee
With these initial conditions, the flow equations for the
ghost and gluon propagators are solved under the constraint of
infrared scaling for the ghost form factor. The resulting full flow of
the ghost dressing function is shown in
Fig. \ref{ghost_flow_full.eps}.

\begin{figure}
\includegraphics[width=0.8\linewidth]{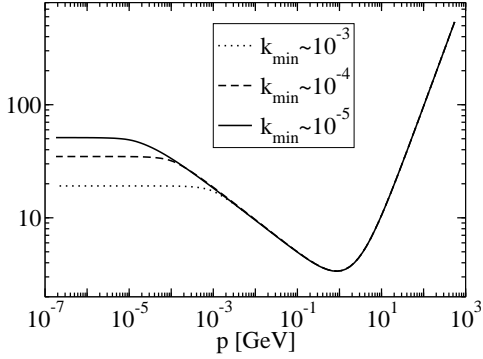}
\caption{Inverse gluon propagator $\omega$ at three minimal cut-off values $k_{min}$.\newline}  
\label{3diff_kmin_omega}
\end{figure}

\begin{figure}
\includegraphics[width=0.8\linewidth]{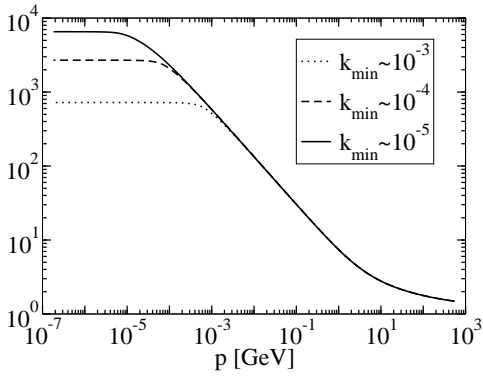} 
\caption{Inverse ghost form factor $d$ at three minimal cut-off values $k_{min}$.}  
\label{3diff_kmin_ghost}
\end{figure}

As the IR cut-off momentum $k$ is
decreased, the ghost form factor $d_k(\vec{p})$ (constant at $k = \Lambda$)
builds up infrared strength and the final solution at $k = k_{min}$ is
shown in Fig. \ref{3diff_kmin_ghost} together with the one for
the gluon energy $\omega_{k_{min}}(\vec{p})$ in Fig. \ref{3diff_kmin_omega}.  It is seen that the IR
exponents, i.e., the slopes of the curves $d_{k_{min}} (\vec{p})$,
$\omega_{k_{min}} (\vec{p})$ do not change as the minimal cut-off $k_{min}$
is lowered. Let us stress that we have assumed infrared scaling of the
ghost form factor but not the horizon condition $d_{k=0}^{-1}(p=0) = 0$. The latter was
obtained from the integration of the flow equation but not put in by
hand (the same is also true for the infrared analysis of the
Dyson--Schwinger equations following from the variational Hamiltonian
approach, i.e., assuming scaling the DSEs yield the horizon
condition).

The infrared exponents extracted from the numerical solutions of the flow equations are
$\alpha = 0{.}28$ and $\beta = 0{.}64$.
They satisfy the sum rule found in \citep{SchLedRei06} resulting from the DSE obtained
in the variational approach but are smaller than the ones of the DSE. Moreover, the
present approach allows to prove the uniqueness of the sum rule (\ref{409}) \citep{Led+10},
analogously to the proof in Landau gauge \citep{FisPaw09}.

Replacing the propagators with running cut-off momentum scale $k$
under the loop integrals of the flow equation by the propagators of
the full theory,
\be\label{32}
d_k (\vec{p}) \to d_{k = 0} (\vec{p}) , \qquad \omega_k (\vec{p}) \to \omega_{k = 0}(\vec{p}) ,
\ee
amounts to taking into account the tadpole diagrams \citep{Led+10}. Then the flow equations can be analytically
integrated and turn into the DSEs obtained in the variational approach 
\citep{FeuRei04}, with explicit UV regularization by
subtraction.\footnote{Instead of the complete right-hand side of Eq.\ 
\eqref{G49} one really obtains only the IR-dominant contribution $\chi^2$.
However, it turns out \citep{Led+10} that the numerical solution of the
equations is the same as in Ref.\ \citep{FeuRei04} over the whole momentum 
range.} This establishes the connection between these two
approaches and highlights the inclusion of a consistent UV
renormalization procedure in the present approach.
The above results encourage further studies, which includes the
flow of the potential between static color sources as well as dynamic
quarks.

%%%%%%%%%%%%%%%%%%%%%%%%%%%%%%%%%%%%%%%%%%%%%%%%
%% BACKMATTER
%%%%%%%%%%%%%%%%%%%%%%%%%%%%%%%%%%%%%%%%%%%%%%%%

\begin{theacknowledgments}
We wish to thank P.~Watson for useful discussions. Financial support
by the DFG under contracts No.\ Re856/6-2,3 and by the Cusanuswerk is greatly acknowledged.
\end{theacknowledgments}

\bibliographystyle{aipproc}
\bibliography{biblio}

\end{document}